\documentclass[12pt]{article}

\usepackage{amsmath,latexsym}

\DeclareMathAlphabet\mathbb  {U}{msb}{m}{n}
\DeclareFontFamily{U}{msb}{} \DeclareFontShape{U}{msb}{m}{n}{
  <5> <6> <7> <8> <9> gen * msbm
  <10> <10.95> <12> <14.4> <17.28> <20.74> <24.88> msbm10
  }{}

\makeatletter
\renewcommand\section{\@startsection{section}{1}{\z@}%
                                     {-3.25ex\@plus -1ex \@minus -.2ex}%
                                     {1.5ex \@plus .2ex}%
                                     {\normalfont\normalsize\bfseries}}
\renewcommand\subsection{\@startsection{subsection}{2}{\z@}%
                                     {-3.25ex\@plus -1ex \@minus -.2ex}%
                                     {1.5ex \@plus .2ex}%
                                     {\normalfont\normalsize\itshape}}

\renewenvironment{thebibliography}[1]
         {\section*{References}\frenchspacing\small
          \begin{list}{[\arabic{enumi}]}
         {\usecounter{enumi}\parsep=2pt\topsep 0pt
         \settowidth{\labelwidth}{[#1]}
         \leftmargin=\labelwidth\advance\leftmargin\labelsep
         \rightmargin=0pt\itemsep=0pt\sloppy}}{\end{list}}

\makeatother

\def\square{\raisebox{-2pt}{$\Box$}}

\begin{document}

\thispagestyle{empty}
\setcounter{page}{0}

$~$\hfill hep-th/0104097
\\
$~$\hfill UWThPh-2001-17
\\
$~$\hfill TUW-01-012

\vskip 20mm

\begin{center}
{\Large\bfseries%
Renormalization of the noncommutative photon 
\\[1ex] self-energy to all orders via Seiberg-Witten map}

\vskip 2cm

{\large
Andreas Bichl$^1$, Jesper Grimstrup$^2$, Harald Grosse$^3$, Lukas
Popp$^4$, \\[1ex] Manfred Schweda$^5$, Raimar Wulkenhaar$^6$}

\vspace{10mm}
       
{\itshape $^{1,2,4,5}$  
Institut f\"ur Theoretische Physik, Technische Universit\"at Wien\\
      Wiedner Hauptstra\ss e 8-10, A-1040 Wien, Austria
\vskip 2ex

$^{3,6}$  Institut f\"ur Theoretische Physik, Universit\"at Wien\\
Boltzmanngasse 5, A-1090 Wien, Austria }

\footnotetext[1]{Work supported in part by 
``Fonds zur F\"orderung der Wissenschaftlichen Forschung'' (FWF) 
under contract P14639-TPH. bichl@hep.itp.tuwien.ac.at}

\footnotetext[2]{Work supported by The Danish Research
  Agency. jesper@hep.itp.tuwien.ac.at} 

\footnotetext[3]{grosse@doppler.thp.univie.ac.at}

\footnotetext[4]{Work supported in part by 
``Fonds zur F\"orderung der Wissenschaftlichen Forschung'' (FWF) 
under contract P13125-PHY. popp@hep.itp.tuwien.ac.at}

\footnotetext[5]{mschweda@tph.tuwien.ac.at}

\footnotetext[6]{Marie-Curie Fellow. raimar@doppler.thp.univie.ac.at}

\vskip 30mm

\begin{abstract}
\noindent
We show that the photon self-energy in quantum electrodynamics on
noncommutative $\mathbb{R}^4$ is renormalizable to all orders (both in
$\theta$ and $\hbar$) when using the Seiberg-Witten map. This is due
to the enormous freedom in the Seiberg-Witten map which represents
field redefinitions and generates all those gauge invariant terms in
the $\theta$-deformed classical action which are necessary to
compensate the divergences coming from loop integrations.
\end{abstract}

\end{center}

\clearpage

\section{Introduction}

Recently Noncommutative Yang-Mills (NCYM) theory has attracted
considerable attention. Partly this is due to its role in string
theory, where NCYM appears as a certain limit in presence of a
constant background field $B$ (see \cite{Seiberg:1999vs} and
references therein). On the other hand, NCYM theory (or better:
Yang-Mills theory on noncommutative $\mathbb{R}^4$) is also an example
of gauge theory on a noncommutative algebra which is interesting on
its own \cite{Connes}. Actually the starting point was a combination
of both \cite{Connes:1998cr}.

Although renormalizable at the one-loop level 
\cite{Martin:1999aq,Sheikh-Jabbari:1999iw,Krajewski:2000ja}, it became
clear that noncommutative field theories suffer from a new type of
infrared divergences \cite{Minwalla:2000px,Matusis:2000jf}
which spoiled renormalization at higher loop
order. Possible problems are ring-type divergences and commutants
\cite{Chepelev:2001hm}. Although this analysis proved
renormalizability for the Wess-Zumino model and complex scalar field
theory \cite{Chepelev:2001hm}, the situation for gauge theory was
desperate. 

An alternative approach to NCYM was proposed by Seiberg and Witten
\cite{Seiberg:1999vs}. They argued from an equivalence of
regularization schemes (point-splitting vs.\ Pauli-Villars) that there
should exist a map (the so-called Seiberg-Witten map) which relates
the noncommutative\footnote{One should better say non-local instead of
  non-commutative because the $\star$-product is a non-local product
  between functions on space-time.}  gauge field $\hat{A}_\mu$ and the
noncommutative gauge parameter $\hat{\lambda}$ to (local) counterparts
$A_\nu$ and $\lambda$ living on ordinary space-time. This approach was
popularized in \cite{Jurco:2000ja} where it was argued that this is
the only way to obtain a finite number of degrees of freedom in
non-Abelian NCYM.

The Seiberg-Witten map leads to a gauge field theory with an infinite
number of vertices and Feynman graphs with unbounded degree of
divergence, which seemed to rule out a perturbative renormalization.
An explicit quantum field theoretical investigation of the
Seiberg-Witten map was first performed in \cite{Bichl:2001nf} for
noncommutative Maxwell theory.  The outcome at one-loop for the photon
self-energy was (to our surprise) gauge invariant and gauge
independent. It was not renormalizable.  However, the divergences were
absorbable by gauge invariant extension terms to the classical action
involving $\theta$ which we interpreted as coming from a more general
scalar product.

It turns out that our extended action is actually a part of the
Seiberg-Witten map when exploiting all its freedom, see also
\cite{Asakawa:1999cu,Jurco:2001rq}. This means that a renormalization
of the Seiberg-Witten map itself is able to remove the one-loop
divergences. This extends to a complete proof of all-order
renormalizability of the photon self-energy. A generalization to other
Green's functions is not obvious, however. This freedom in the
Seiberg-Witten map can be regarded as a field redefinition.

\section{The freedom in the Seiberg-Witten map}

We consider NCYM theory with fermions, regarded as a model
on ordinary Minkowski space (with metric $g^{\mu\nu}$), subject to the
altered (non-local) multiplication law for functions $f,g$ on
space-time:
\begin{align}
\label{Moyal}
(f \star g)(x) = \int d^4y \,d^4z \; 
\delta^4(y-x) \,\delta^4(z-x)\;
\exp\Big(\mathrm{i} \theta^{\mu\nu} \frac{\partial}{\partial y^\mu} 
\frac{\partial}{\partial z^\nu} \Big) \big( f(y) g(z)\big) ~.
\end{align}
The real parameter $\theta_{\mu\nu}=-\theta_{\nu\mu}$ will be regarded
as a constant external field of power-counting dimension $-2$. 

The Seiberg-Witten map \cite{Seiberg:1999vs} expresses the
noncommutative gauge fields $\hat{A}_\mu{=}\hat{A}_\mu[A_\nu,\theta]$,
the infinitesimal gauge parameter
$\hat{\lambda}{=}\hat{\lambda}[\lambda,A_\nu,\theta]$ and the fermions
$\hat{\psi}{=}\hat{\psi}[\psi,A_\nu,\theta]$,
$\hat{\bar{\psi}}{=}\hat{\bar{\psi}}[\bar{\psi},A_\nu,\theta]$, which
are multiplied according to (\ref{Moyal}), as formal
power series of the corresponding gauge-equivalent commutative
(but non-Abelian) objects $A_\mu,\lambda,\psi,\bar{\psi}$ to be multiplied
in the ordinary way. The gauge-equivalence condition is
\begin{align}
\delta_{\hat{\lambda}} \hat{A}_\mu = 
\delta_{\lambda} \hat{A}_\mu ~,\qquad
\delta_{\hat{\lambda}} \hat{\psi} = 
\delta_{\lambda} \hat{\psi} ~,\qquad
\delta_{\hat{\lambda}} \hat{\bar{\psi}} = 
\delta_{\lambda} \hat{\bar{\psi}} ~,
\label{ge}
\end{align}
with initial condition
\begin{align}
\label{init}
\hat{A}_\mu[A_\nu,\theta{=}0]=A_\mu\,,\quad
\hat{\lambda}[\lambda,A_\nu,\theta{=}0]=\lambda\,,\quad
\hat{\psi}[\psi,A_\nu,\theta{=}0]=\psi \,,\quad 
\hat{\bar{\psi}}[\bar{\psi},A_\nu,\theta{=}0]=\bar{\psi}\,.
\end{align}
The noncommutative gauge transformations are defined by
\begin{align}
\delta_{\hat{\lambda}} \Gamma = \int d^4x\;\Big( & \mathrm{tr}\Big(
(\partial_\mu \hat{\lambda} 
- \mathrm{i} (\hat{A}_\mu \star \hat{\lambda} 
- \hat{\lambda} \star \hat{A}_\mu))\star 
 \frac{\delta \Gamma}{\delta \hat{A}_\mu}\Big)
\nonumber
\\
& + \Big\langle \frac{\Gamma \overleftarrow{\delta}}{\delta
  \hat{\psi}} \star 
(\mathrm{i} \hat{\lambda} \star \hat{\psi}) \Big\rangle
+ \Big\langle (- \mathrm{i} \hat{\bar{\psi}} \star \hat{\lambda}) \star
\frac{\overrightarrow{\delta}\Gamma}{\delta \hat{\bar{\psi}}} \Big\rangle
\Big)
\end{align}
and the commutative\footnote{Although we are first of all interested
  in QED, we present everything as far as possible in a way
  which also applies to $\theta$-deformed Yang-Mills theory.} ones by
\begin{align}
\delta_{\lambda} \Gamma = \int \!\! d^4 x\;\Big(\mathrm{tr} \Big(
(\partial_\mu \lambda 
- \mathrm{i} (A_\mu \lambda - \lambda A_\mu))
 \frac{\delta \Gamma}{\delta A_\mu}\Big)
+ \Big\langle \frac{\Gamma \overleftarrow{\delta}}{\delta \psi} 
(\mathrm{i}  \lambda \psi)\Big\rangle 
+ \Big\langle (- \mathrm{i} \bar{\psi} \lambda) 
\frac{\overrightarrow{\delta}\Gamma}{\delta \bar{\psi}} 
\Big\rangle \Big)\;.
\label{gt}
\end{align}
The bracket $\langle~\rangle$ means trace in colour and spinor space.

As shown in \cite{Asakawa:1999cu} there is a big variety of solutions
of (\ref{ge}),(\ref{init}) correspondinging to field redefinitions.
Here we take a subclass of the solutions derived in
\cite{Asakawa:1999cu}\footnote{Similar ideas are used in
  \cite{Jurco:2001rq} where a general formalism for the construction
  of the Seiberg-Witten map is given.}. We denote by
$\hat{A}^{(n)}_\mu$ a solution of (\ref{ge}),(\ref{init}) up to order
$n$ in $\theta$. Then, a further solution up to the same order $n$ is
obtained by adding any gauge-covariant term with
\emph{exactly}\footnote{This is important: $\mathbb{A}_\mu^{(n)}$
  contains exactly $n$ factors of $\theta$ whereas $\hat{A}_\mu^{(n)}$
  contains $0\leq j \leq n$ factors of $\theta$.}  $n$ factors of
$\theta$,
\begin{align}
\hat{A}_\mu^{(n)}{}'& =\hat{A}_\mu^{(n)} + 
\mathbb{A}_\mu^{(n)}~, \nonumber
\\
\mathbb{A}_\mu^{(n)} &= \sum_{(i)} \kappa_i^{(n)} \big(
\underbrace{g^{**} \cdots g^{**}}_{2n}\, 
\underbrace{\theta_{**}\cdots \theta_{**}\,}_n \,
\underbrace{D_{*} \cdots D_{*}\;}_{l_1} (F_{**}) \cdots 
\underbrace{D_{*} \cdots D_{*}\;}_{l_k} (F_{**}) \big)^{(i)}_\mu~,
\label{C}
\end{align}
where $\sum_{j=1}^k l_j=2n{+}1{-}2k$. This condition guarantees that
$\hat{A}_\mu^{(n)}{}'$ has the same power-counting
dimension\footnote{Power-counting dimensions $\mathrm{dim}$ are
defined as follows: $\mathrm{dim}(A_\mu) =  \mathrm{dim}(\hat{A}_\mu)=
  1$, $\mathrm{dim}(\psi)=  \mathrm{dim}(\hat{\psi})= 
\mathrm{dim}(\bar{\psi})=  \mathrm{dim}(\hat{\bar{\psi}})=
\frac{3}{2}$, 
$\mathrm{dim}(\partial_\mu)=1$, $\mathrm{dim}(m)=1$,
$\mathrm{dim}\big(\int d^4x\big)=-4$, 
$\mathrm{dim}(\delta^4(x{-}y))=4$, 
$\mathrm{dim}\big(\int d^4p\big)=4$, $\mathrm{dim}(\delta^4(p{-}q))=-4$, 
$\mathrm{dim}(\theta)=-2$.
  \label{pow}} (=1) as
$A_\mu$ when taking $\theta$ of power-counting dimension $-2$.  Each
$*$ in (\ref{C}) stands for a Lorentz index (all but the free lower
index $\mu$ are summation indices).  $D_\nu=\partial_\nu -\mathrm{i}
[A_\nu,\,.\,]$ is the covariant derivative and
$F_{\mu\nu}=\partial_\mu A_\nu - \partial_\nu A_\mu - \mathrm{i}
[A_\mu, A_\nu]$ the (commutative) Yang-Mills field strength.  The sum
is over all index structures $(i)$ and $\kappa_i^{(n)}\in \mathbb{R}$
is a free parameter. Inserted into the gauge-equivalence (\ref{ge})
there is on the l.h.s.\ at order $n$ no further factor of $\theta$
coming from $\hat{\lambda}$ or the $\star$-product possible:
\begin{align}
  \label{ginv}
\delta_{\hat{\lambda}} \hat{A}_\mu^{(n)}{}' = 
\delta_{\hat{\lambda}} \hat{A}_\mu^{(n)} -\mathrm{i} [
\mathbb{A}_\mu^{(n)},\lambda ] \equiv \delta_{\lambda} \hat{A}_\mu^{(n)}{}'
\qquad
\text{up to order $n$}\,.  
\end{align}
Thus, $\hat{A}_\mu^{(n)}{}'$ is a solution of the gauge-equivalence
condition if $\hat{A}_\mu^{(n)}$ is, up to order $n$. 
The effect of $\mathbb{A}_\mu^{(n)}$ on the noncommutative field
strength $\hat{F}_{\mu\nu} = \partial_\mu \hat{A}_\nu - \partial_\nu
\hat{A}_\mu - \mathrm{i}(\hat{A}_\mu \star \hat{A}_\nu - \hat{A}_\nu
\star \hat{A}_\mu )$ is up to order $n$ given by
\begin{align}
\hat{F}_{\mu\nu}^{(n)}{}' = \hat{F}_{\mu\nu}^{(n)} + D_\mu
\mathbb{A}^{(n)}_\nu- D_\nu \mathbb{A}^{(n)}_\mu~,
\end{align}
because no factor $\theta$ from $\hat{A}_\alpha$ or the
$\star$-product can be combined with $\mathbb{A}^{(n)}_\mu$ up to
order $n$. The noncommutative Yang-Mills action is
\begin{align}
\label{SB}
\hat{\Sigma} =
-\frac{1}{4g^2} \int d^4x \;\mathrm{tr}(\hat{F}^{\mu\nu} \hat{F}_{\mu\nu})~.
\end{align}
Defining $\hat{\Sigma}^{(n)}{}'$ as the result of (\ref{SB}) when
replacing $\hat{F}_{\mu\nu}^{(n)}$ by $\hat{F}_{\mu\nu}^{(n)}{}'$ and
the commutative actions $\Sigma^{(n)}{}'$ and $\Sigma^{(n)}$ as the
Seiberg-Witten map of $\hat{\Sigma}^{(n)}{}'$ and
$\hat{\Sigma}^{(n)}$, we obtain up to order $n$ in $\theta$
\begin{align}
\label{YM}
\Sigma^{(n)}{}' =\Sigma^{(n)}
+\frac{1}{g^2}\int d^4x\; \mathrm{tr}\Big(F^{\mu\nu} D_\nu 
\mathbb{A}^{(n)}_\mu \Big)
=\Sigma^{(n)}
+\frac{1}{g^2}\int d^4x\; \mathrm{tr}\Big(
(D_\nu F^{\nu\mu}) \mathbb{A}^{(n)}_\mu \Big)~.
\end{align}
The part $\Sigma^{(n)}{}'-\Sigma^{(n)}$ of the action represents due
to (\ref{C}) and the dimension assignment in footnote \ref{pow} a
gauge invariant action of power-counting dimension 0 with $n$ factors
of $\theta$.  Gauge invariance means that application of the operator
$\delta_\lambda$ defined in (\ref{gt}) yields zero.  The action
$\Sigma^{(n)}{}$ is gauge invariant at any order $k\leq n$ in
$\theta$, thus yielding at order $n$ in $\theta$ terms which are also
present in $\Sigma^{(n)}{}'-\Sigma^{(n)}$. These terms in
$\Sigma^{(n)}$ can be regarded as a shift to $\kappa^{(n)}_i$.

Now we pass to quantum field theory and compute Feynman graphs. The
loop integrations will produce divergent 1PI-Green's functions which
under the assumption of an invariant renormalization
scheme\footnote{If no invariant renormalization scheme is available
  (or if one chooses a non-invariant scheme for some reason) one
  should attempt to restore gauge invariance via the quantum action
  principle and a parameter redefinition. Gauge anomalies are an
  obstruction to such a program.}  are gauge invariant field
polynomials of power-counting dimension 0. We hope to remove all of
these divergences with $n$ factors of $\theta$ by a
$\hbar$-redefinition of $\kappa^{(n)}_i$. The problem is that
(\ref{YM}) generates only a subset of all possible gauge invariant
actions. For the photon self-energy in $\theta$-deformed QED we are
able to show that all divergences actually belong to this subset
(Section \ref{ps}). Before we will address the question of a
physical meaning of the $\kappa_i^{(n)}$.

\section{Field redefinitions}
\label{fr}

It is possible to rewrite (\ref{YM}) in the
following form:
\begin{align}
  \label{redef}
\Sigma^{(n)}{}' &= \Sigma^{(n)} +\delta_{\mathbb{A}}^{(n)} \Sigma^{(n)}
\qquad \text{up to order $n$}~,
\nonumber
\\
\delta_{\mathbb{A}}^{(n)} \Gamma &= 
\int d^4x\, \mathbb{A}^{a\,(n)}_\mu(x)
\frac{\delta \Gamma}{\delta A_\mu^a(x)} ~,
\end{align}
where $\Gamma$ is any functional depending on $A^a_\mu$. In
(\ref{redef}) we use now the component formulation induced by
$A_\mu=A^a_\mu T_a$, with $[T_a,T_b]=\mathrm{i} f_{ab}^{~~c} \,T_c$.
This suggests to consider $\mathbb{A}^{a\,(n)}_\mu$ as a field
redefinition of $A_\mu$. As such we must check how it commutes with
the local Ward identity operator with respect to a variation of the
gauge field,
\begin{align}
 \label{W}
W^\lambda_a(y) = \frac{\delta}{\delta \lambda^a(y)} \delta_\lambda
= -\partial^y_\mu \frac{\delta}{\delta A^a_\mu(y)} - f_{ab}^{~~c} 
A^b_\mu(y) \frac{\delta}{\delta A^c_\mu(y)}~,
\end{align}
where $\delta_\lambda$ is defined in (\ref{gt}). To the commutator
$\delta_{\mathbb{A}}^{(n)} W^\lambda_a(y) \Gamma - 
W^\lambda_a (y) \delta_{\mathbb{A}}^{(n)} \Gamma $ there is only a
contribution if both operators $\delta_{\mathbb{A}}^{(n)}$ and
$W^\lambda_a(y)$ hit the same field $A_\mu$ in $\Gamma$, hence it is
sufficient to consider $\Gamma \mapsto A_\mu^c(x)$. Then we have
\[
\delta_{\mathbb{A}}^{(n)} W^\lambda_a(y) A_\mu^c(x) = 
- f_{ab}^{~~c} \mathbb{A}_\mu^{b\,(n)}(y) \, \delta(x-y)
\]
due to (\ref{W}),(\ref{redef}) and 
\[
W^\lambda_a(y) \delta_{\mathbb{A}}^{(n)} A_\mu^c(x) = 
W^\lambda_ a(y) \mathbb{A}^{c\,(n)}_\mu(x) = 
- f_{ab}^{~~c} \mathbb{A}_\mu^{b\,(n)}(y) \,\delta(x-y)
\]
because of the covariance 
$\delta_\lambda \mathbb{A}_\mu^{(n)} =
\mathrm{i}[\lambda,\mathbb{A}_\mu^{(n)}]$, see (\ref{C}). This means 
\begin{align}
[\delta_{\mathbb{A}}^{(n)}, W^\lambda_a(y)] \equiv 0~,
\end{align}
i.e.\ all $\kappa^{(n)}_i$ in $\mathbb{A}_\mu$ must be regarded as
parametrizations of field redefinitions.

\section{Quantum field theory}

The basic object in quantum field theory is the generating functional
$\Gamma$ of one-particle irreducible (1PI) Green's functions (with $n$
factors of $\theta$)
\begin{align}
  \label{GF}
\Gamma[A_{c\ell}]^{(n)} = 
\sum_{N \geq 2} \frac{1}{N!} 
\int d^4x_1\dots d^4x_N \; A_{\mu_1\,c\ell}^{a_1}(x_1) \cdots 
A_{\mu_N\,c\ell}^{a_N}(x_N) \;\langle 0| T A^{\mu_1}_{a_1}(x_1)
\dots A^{\mu_N}_{a_N}(x_N)|0\rangle_{\mathrm{1PI}}^{(n)}
\end{align}
in terms of classical fields $A_{c\ell}$. Colour indices are
denoted by $a_i$. The vacuum expectation value of the time-ordered
product of fields in (\ref{GF}) is the Fourier transform of the
$N$-point vertex functional in momentum space,
\begin{align}
&\langle 0| T A^{\mu_1}_{a_1}(x_1)
\dots A^{\mu_N}_{a_N}(x_N)|0\rangle_{\mathrm{1PI}}^{(n)}
\label{verf}
\nonumber
\\
&\qquad = \int \frac{d^4p_1}{(2\pi)^4} \dots 
\frac{d^4p_N}{(2\pi)^4} \;\delta^4(p_1+\dots+p_N)\;
\mathrm{e}^{-\mathrm{i}p_1x_1} \cdots 
\mathrm{e}^{-\mathrm{i}p_Nx_N} \;
\Gamma^{\mu_1\dots\mu_N\,(n)}_{a_1\dots a_N}(p_1,\dots p_N)
\end{align}
with $\mathrm{dim}\big(\Gamma^{\mu_1\dots\mu_N}_{a_1\dots
  a_N}(p_1,\dots p_N)\big) = 4{-}N$. Due to the $n$ factors of
$\theta$, the momentum space degree of divergence of
$\Gamma^{\mu_1\dots\mu_N\,(n)}_{a_1\dots a_N}(p_1,\dots p_N)$ is
$\omega=4+2n-N$. The local Ward identity operator (\ref{W}) applied
to (\ref{GF}),
\begin{align}
  \label{loW}
W_a(y) \Gamma[A_{c\ell}]^{(n)}= -\partial_\rho^y 
\frac{\delta \Gamma[A_{c\ell}]^{(n)}}{\delta 
A_{\rho\,c\ell}^a(y)} 
- f^{~~c}_{ab}\; A^b_{\rho\,c\ell}(y) 
\frac{\delta \Gamma[A_{c\ell}]^{(n)}}{\delta 
A_{\rho\,c\ell}^c(y)}~.
\end{align}
%where $f^{~~c}_{ab}$ are the structure constants of the Lie algebra
%defined by $[X,Y]_a=\mathrm{i} f^{~~c}_{ab} X^b Y_c$. 
is evaluated in presence of an invariant renormalization scheme to
\[
W_a(y) \Gamma[A_{c\ell}]^{(n)} = \partial^\mu\partial_\mu B(y)~.
\]
Here, $B$ is the multiplier field required for gauge-fixing. In a
linear gauge\footnote{We refer to \cite{Bichl:2001nf} for a natural
  nonlinear gauge in $\theta$-deformed Maxwell theory.} there are no
vertices with external $B$-lines and thus no divergent 1PI Green's
functions with external $B$ (furthermore, $B$ is independent of
$\theta$). Therefore we have the local Ward identity $W_a(y)
\Gamma[A_{c\ell}]^{(n)}=0$ for $\Gamma[A_{c\ell}]^{(n)}$ being 1PI and
divergent. Then, functional derivation of (\ref{loW}) with respect to
$A_{\mu_1\,c\ell}^{a_1}(x_1) \dots A_{\mu_N\,c\ell}^{a_N}(x_N)$,
followed by putting the remaining $A_{\nu\,c\ell}^{b}(z)=0$, gives
\begin{align}
  \label{WI}
0 &= - \partial_\rho^y \langle0|T A^\rho_a (y) A^{\mu_1}_{a_1} (x_1) 
\dots A^{\mu_N}_{a_N} (x_N) |0\rangle_\mathrm{1PI}^{(n)}
\\
& - \sum_{j=1}^N  f^{~~c}_{aa_j} \,\delta(y-x_j) \;
\langle0|T A^{\mu_j}_c (y) A^{\mu_1}_{a_1} (x_1) 
\dots A^{\mu_{j-1}}_{a_{j-1}} (x_{j-1}) 
A^{\mu_{j+1}}_{a_{j+1}} (x_{j+1}) \dots
A^{\mu_N}_{a_N} (x_N) |0\rangle_\mathrm{1PI}^{(n)}~.
\nonumber
\end{align}

\section{The photon self-energy in $\theta$-deformed QED}
\label{ps}

We recall that $\omega=4+2n-N$ is the power-counting degree of
divergence for the $N$-point photon vertex functionals with $n$
factors of $\theta$, independent of the internal structure of the
Feynman graphs. Due to translation invariance (or momentum
conservation) we therefore have
\begin{align}
\langle0|T A^{\mu_1}_{a_1} (x_1) &
\dots  A^{\mu_N}_{a_N} (x_N) |0\rangle_\mathrm{1PI}^{(n)} 
\label{Delta}
\\
\nonumber
&= \sum_{(i)} \kappa_i'
\Big(\underbrace{g^{**} \cdots g^{**}}_{2n-N+2}\;
\underbrace{\theta_{**} \cdots \theta_{**}\,}_{n}\;
\underbrace{\partial_* \dots \partial_*\,}_{4+2n-N} \big(
\delta(x_1{-}x_2) \cdots \delta(x_{N-1} {-} x_N) \big)
\Big)^{(i)\,\mu_1\dots\mu_N}
\end{align}
The sum is over all index structures $(i)$ with appropriate numerical
factors $\kappa_i$, and the derivatives are with respect to any of the
coordinates $x_1,\dots,x_N$. We insert (\ref{Delta}) into (\ref{GF})
and integrate by parts. Assuming an invariant renormalization scheme
(such as dimensional regularization), the local Ward identity
(\ref{WI}), with $f_{ab}^{~~c}=0$, implies that the generating
functional $\Gamma[A_{c\ell}]^{(n)}$ must be a function of the
classical field strength $F_{c\ell\,\mu\nu} = \partial_\mu
A_{c\ell\,\nu}- \partial_\nu A_{c\ell\,\mu}$:
\begin{align}
\Gamma[A_{c\ell}]^{(n)} & = \sum_{N \geq 2} \frac{1}{N!}
\int \! d^4x_1\dots d^4x_N \sum_{(i)} \kappa_i 
\Big(\underbrace{g^{**} \cdots g^{**}}_{2n+2}\;
\underbrace{\theta_{**} \cdots \theta_{**}\,}_{n}\; 
 \nonumber
\\
& \times 
\underbrace{\partial_* \dots \partial_*\,}_{4+2n-2N} 
\big(F_{c\ell\,**}(x_1) \dots F_{c\ell\,**}(x_N)\big)\,
\delta(x_1{-}x_2) \cdots \delta(x_{N-1} {-} x_N) \Big)^{(i)}.
\label{invG}
\end{align}
{}From the Ward identity it follows in particular that $N\leq n+2$. 

Now we specialize (\ref{invG}) to the photon self-energy, i.e.\ to the
$N=2$ part in (\ref{invG}). All derivatives can be assumed acting on
$F_{c\ell\,**}(x_1)$. There are both $2n$ indices on $\theta_{**}$ and
$\partial_*$, but we have $\theta^{\alpha\beta}\partial_\alpha^{x_1}
\partial_\beta^{x_1}=0$. Therefore, there is for $n\geq 1$ 
always one of the terms
\[
\partial^\mu F_{\mu\nu}\qquad \text{or} \qquad  
\partial^\alpha \partial_\alpha F_{\mu\nu}
=\partial_\mu  \partial^\alpha F_{\alpha\nu} 
- \partial_\nu \partial^\alpha F_{\alpha\mu} 
\]
in the $N=2$ part of (\ref{invG}). But this is according to (\ref{YM})
nothing but the structure of a noncommutative Maxwell action after
Seiberg-Witten map (with $D_\nu\equiv \partial_\nu$), which thus is
able to absorb all divergences coming from loop integrations: The
two-point function in the noncommutative Maxwell action
\begin{align}
\label{Sig}
\hat{\Sigma}' = -\frac{1}{4 g^2} \int d^4x \;
\hat{F}^{\mu\nu}{}'\hat{F}_{\mu\nu}'
\end{align}
is renormalizable at order $n$ in $\theta$ and any order $L$ in
$\hbar$ due to the gauge-covariant terms $\mathbb{A}^{(n)}_\mu$ in the
Seiberg-Witten map, i.e.\ by a $\hbar$-redefinition of
$\kappa_i^{(n)}$ which preserves the form of (\ref{Sig}).

The argument does not wok for $N$-point functions with $N\geq 3$. For
instance, it is now possible to contract all derivatives in
(\ref{invG}) with the factors of $\theta$ as the following
contribution to the $3$-point function shows:
\[
\int d^4x_1\,d^4x_2\,d^4x_3\;
\theta^{\gamma\delta} \,
\Big(\prod_{i=1}^3  \theta^{\alpha_i\beta_i} \partial_{\alpha_i}^{x_1} 
\partial_{\beta_i}^{x_2}\Big) \Big(
F_{\gamma\delta}(x_1) F^{\mu\nu}(x_2) F_{\mu\nu}(x_3) \Big)
\, \delta(x_1-x_2)\delta(x_2-x_3)~.
\]
The complete renormalization of NCYM theories remains an open problem.

\subsection{One-loop photon self-energy at second order in $\theta$}

As an example let us look at the lowest orders of noncommutative
Maxwell theory studied in \cite{Bichl:2001nf}. In
order $\theta^1$ there is only one\footnote{The free index $\mu$ can
  not occur via $\partial_\mu$ because this would lead to a vanishing
  field strength. Moreover, one has to take the Bianchi identity into consideration.}
gauge covariant (here: invariant)
extension to the Seiberg-Witten map:
\[
\mathbb{A}_\mu^{(1)}=\kappa_1^{(1)} \theta_{\mu\alpha} \partial_\beta
F^{\alpha\beta}
\]
which, however, drops out of the Maxwell action,
$F^{\mu\nu} \theta_{\mu\alpha} \partial_\nu \partial_\beta
F^{\alpha\beta}= -  \theta_{\mu\alpha} (\partial_\nu F^{\mu\nu})
(\partial_\beta F^{\alpha\beta}) =0$. 
At order $\theta^2$ we have, up to total derivatives
$\partial_\mu(\,.\,)$ and Bianchi identity, four different
terms\footnote{There are no divergent graphs of order 2 in $\theta$
  with more than two external photon lines.}  in (\ref{C}):
\begin{align}
\mathbb{A}_\mu^{(2)}=
& \Big(
\kappa_1^{(2)} g^{\alpha\gamma} g^{\beta\delta}
g^{\lambda\rho} g^{\sigma\tau}
\theta_{\alpha\beta} \theta_{\gamma\delta}
\partial_\lambda \partial_\rho \partial_\sigma
F_{\tau\mu}
+
\kappa_2^{(2)} g^{\alpha\gamma} g^{\beta\lambda}
g^{\delta\rho} g^{\sigma\tau}
\theta_{\alpha\beta} \theta_{\gamma\delta}
\partial_\lambda \partial_\rho \partial_\sigma
F_{\tau\mu}
\nonumber
\\
& +
\kappa_3^{(2)} g^{\beta\sigma} g^{\gamma\tau} 
g^{\alpha\lambda} g^{\delta\rho}
\theta_{\mu\beta} \theta_{\gamma\delta}
\partial_\alpha \partial_\lambda \partial_\rho 
F_{\sigma\tau}
+ 
\kappa_4^{(2)} g^{\gamma\tau} g^{\beta\delta}
g^{\alpha\lambda} g^{\rho\sigma}
\theta_{\mu\beta} \theta_{\gamma\delta}
\partial_\alpha \partial_\lambda \partial_\rho 
F_{\sigma\tau}
\Big)\,.
\label{ka2}
\end{align}
These lead to the following terms in the action (\ref{YM}):
\begin{align}
\Sigma^{(2)}{}' =\Sigma^{(2)} + \frac{1}{g^2} \int d^4x\;A_\mu & \Big(
(g^{\mu\nu} \square -\partial^\mu\partial^\nu) 
(\kappa_1^{(2)} \theta^2 \square^2
+ \kappa_2^{(2)} \tilde{\tilde{\square}} \square)
+ \kappa_3^{(2)} \tilde{\partial}^\mu \tilde{\partial}^\nu \square^2
\nonumber
\\
&
+ \kappa_4^{(2)} (\theta^{\mu\alpha} \theta^\nu_{~\alpha} \square^3
+ (\tilde{\tilde{\partial}}{}^\mu \partial^\nu
+ \tilde{\tilde{\partial}}{}^\nu \partial^\mu )\square^2 
+ \partial^\mu \partial^\nu \tilde{\tilde{\square}} \square)
\Big) A_\nu~,
  \label{act}
\end{align}
where $\square=\partial^\alpha\partial_\alpha$,
$\tilde{\partial}^\alpha=\theta^{\alpha\beta} \partial_\beta$,
$\tilde{\tilde{\partial}}{}^\alpha=\theta^{\alpha\beta}
\tilde{\partial}_\beta$,
$\tilde{\tilde{\square}}=\tilde{\partial}^\alpha
\tilde{\partial}_\alpha$ and $\theta^2=\theta^{\alpha\beta}
\theta_{\alpha\beta}$. The rhs of (\ref{act}) can now be rewritten in
the following form:
\begin{align}
\frac{1}{g^2} \int d^4 x \;\partial_\rho F^{\rho\mu}(x)
\mathbb{A}^{(2)}_\mu(x) 
&=  \frac{1}{2 g^2} \int d^4 x\,d^4y \; 
A_\mu(x) A_\nu(y)\; \langle0| T A^\mu(x) A^\nu(y)
|0\rangle^{(2)}_\mathrm{1PI}~,
\quad\text{with}
\nonumber
\\[1ex]
\langle0| T A^\mu(x) A^\nu(y)|0\rangle^{(2)}_\mathrm{1PI}
& = \Big(
(g^{\mu\nu} \square -\partial^\mu\partial^\nu) 
(2\kappa_1^{(2)} \theta^2 \square^2
+ 2\kappa_2^{(2)} \tilde{\tilde{\square}} \square)
+ 2\kappa_3^{(2)} \tilde{\partial}^\mu \tilde{\partial}^\nu \square^2
\nonumber
\\
&\quad
+ 2\kappa_4^{(2)} (\theta^{\mu\alpha} \theta^\nu_{~\alpha} \square^3
+ (\tilde{\tilde{\partial}}{}^\mu \partial^\nu
+ \tilde{\tilde{\partial}}{}^\nu \partial^\mu )\square^2 
+ \partial^\mu \partial^\nu \tilde{\tilde{\square}} \square)
\Big)_x \delta(x-y)~. 
\label{twopoint}
\end{align}
Comparing (\ref{act}) with the one-loop
calculation in \cite{Bichl:2001nf} we see that the following
renormalization of $\kappa^{(2)}_1 ,\dots,\kappa^{(2)}_4$,
\begin{align}
\kappa^{(2)}_1 & \mapsto  \kappa^{(2)}_1 - \frac{g^2
  \hbar}{16 (4\pi)^2\varepsilon} ~, &
\kappa^{(2)}_2 & \mapsto  \kappa^{(2)}_2 + \frac{g^2
  \hbar}{20 (4\pi)^2\varepsilon} ~,
\nonumber
\\
\kappa^{(2)}_3 & \mapsto  \kappa^{(2)}_3 + \frac{g^2
  \hbar}{60 (4\pi)^2\varepsilon} ~, &
\kappa^{(2)}_4 & \mapsto  \kappa^{(2)}_4 + \frac{g^2
  \hbar}{8 (4\pi)^2\varepsilon} ~, 
  \label{renorm}
\end{align}
cancels precisely the one-loop divergences in the photon self-energy.
In other words, (\ref{renorm}) provides a formal power series
$\kappa^{(2)}_i [\hbar]$ such that the one-loop photon self-energy
Greens's function is at order $\theta^2$ renormalizable. However, 
(\ref{renorm}) represent unphysical renormalizations because the
$\kappa$'s parametrize field redefinitions, see Section \ref{fr}. This
means that at order $0$ in $\hbar$ the $\kappa^{(2)}_i$ may be set to zero.

\section{Extension to any order in $\theta$}

It remains to prove that the gauge-equivalence (\ref{ge}) of the
Seiberg-Witten map can be extended to order $n{+}1$ in $\theta$. This
is not clear a priori because the gauge transformations
$\delta_{\hat{\lambda}}$ and $\delta_\lambda$ applied to $\mathbb{A}^{(n)}_\mu$
produce very different results at higher order in $\theta$. 

We expand $\hat{A}^{(n+1)}_\mu$ into a Taylor series:
\begin{align}
\hat{A}^{(n+1)}_\mu &= \sum_{k=0}^{n+1} \frac{1}{k!}
\theta^{\alpha_1\beta_1} \cdots \theta^{\alpha_k\beta_k} 
\Big(\frac{\partial^k \hat{A}^{(n+1)}_\mu}{\partial
  \theta^{\alpha_1\beta_1} \dots
  \theta^{\alpha_k\beta_k}}\Big)_{\theta=0} 
\nonumber
\\
  \label{taylor}
& = A_\mu + \sum_{k=1}^{n+1} \frac{1}{k!}
\theta^{\alpha\beta} \theta^{\alpha_2\beta_2} 
\cdots \theta^{\alpha_k\beta_k} 
\Big(\frac{\partial^{k-1}}{\partial 
  \theta^{\alpha_2\beta_2} \dots
  \theta^{\alpha_k\beta_k}}
\Big(\frac{\partial \hat{A}^{(n+1)}_\mu}{\partial \theta^{\alpha\beta}}
\Big)\Big)_{\theta=0} ~.
\end{align}
We recall now the Seiberg-Witten differential equation\footnote{We
  would like to stress that (\ref{dgl}) guarantees 
$\mathrm{dim}(\hat{A}_\mu)=1$ to all
  orders of $\theta$.} \cite{Seiberg:1999vs}
\begin{align}
\label{dgl}
\frac{\partial \hat{A}_\mu}{\partial \theta^{\alpha\beta}}
= 
-\frac{1}{8}\Big\{
\hat{A}_\alpha , (\hat{F}_{\beta\mu}+\partial_\beta \hat{A}_\mu)
\Big\}_\star
+\frac{1}{8}\Big\{
\hat{A}_\beta , (\hat{F}_{\alpha\mu}+\partial_\alpha \hat{A}_\mu) \Big\}_\star
\end{align}
for a solution $\hat{A}_\mu$ of (\ref{ge}), where $\{X,Y\}_\star :=
X\star Y + Y\star X$ is the $\star$-anticommutator. We see that
$\frac{\partial \hat{A}^{(n+1)}_\mu}{\partial \theta^{\alpha\beta}}$
requires knowledge of only $\hat{A}^{(n)}_\nu$ (i.e.\ of the
Seiberg-Witten map up to order $n$). Taking the general
order-$n$ solution (\ref{C}), i.e.\ including $\mathbb{A}^{(n)}_\nu$,
we obtain a Seiberg-Witten map up to order $n{+}1$,
\begin{align}
\hat{A}^{(n+1)}_\mu{}' & = A_\mu -\frac{1}{4} 
\sum_{k=1}^{n+1} \frac{1}{k!}
\theta^{\alpha_1\beta_1} 
\cdots \theta^{\alpha_k\beta_k} 
\Big(\frac{\partial^{k-1}}{\partial 
  \theta^{\alpha_2\beta_2} \dots
  \theta^{\alpha_k\beta_k}}
\Big\{
\hat{A}^{(n)}_{\alpha_1}{}', (\hat{F}^{(n)}_{\beta_1\mu}{}'
+\partial_{\beta_1} \hat{A}^{(n)}_\mu{}') \Big\}_\star
\Big)_{\theta=0} 
\nonumber
\\
  \label{n+1}
& + \mathbb{A}^{(n+1)}_\mu~,
\end{align}
which implies renormalizability up to order $n{+}1$ in
$\theta$. Accordingly, the noncommutative gauge parameter is at order
$n{+}1$ in $\theta$ obtained as
\begin{align*}
\hat{\lambda}^{(n+1)} & = \lambda - \frac{1}{4} 
\sum_{k=1}^{n+1} \frac{1}{k!}
\theta^{\alpha_1\beta_1} 
\cdots \theta^{\alpha_k\beta_k} 
\Big(\frac{\partial^{k-1}}{\partial 
  \theta^{\alpha_2\beta_2} \dots
  \theta^{\alpha_k\beta_k}}
\Big\{
\hat{A}^{(n)}_{\alpha_1}{}', \partial_{\beta_1} \hat{\lambda}^{(n)}) 
\Big\}_\star \Big)_{\theta=0} ~.
\end{align*}
Thus we have proved by induction that the photon self-energy arising
from the noncommutative Maxwell action (\ref{Sig}) is (under the
assumption of an invariant renormalization scheme) renormalizable
to all orders in $\theta$ and $\hbar$ via a general Seiberg-Witten
map. Observe that $\hat{A}^{(n+1)}_\mu{}'$ is a complicated nonlinear
function of $\kappa_i^{(j)}$ for $j\leq n$.

\section{Remarks on the fermionic action}

We would like to extend the renormalizability proof for the photon
self-energy to Green's functions in $\theta$-deformed QED
\cite{Bichl:2001gu} containing fermions. So far we did not succeed,
nevertheless we present some ideas which hopefully turn out to be
useful. On that level we can formulate everything for Yang-Mills
theory with fermions.

In analogy to (\ref{C}) we add to a solution $\hat{\psi}^{(n)}$ of the
gauge-equivalence (\ref{ge}) the most general gauge-covariant term in
$\psi$ with exactly $n$ factors of $\theta$:
\begin{align}
\hat{\psi}^{(n)}{}' & =\hat{\psi}^{(n)} +\Psi^{(n)}~,\qquad
\nonumber
\\
\Psi^{(n)} & = \sum_{(i)}  \tilde{\kappa}^{(n)}_i \Big( m^t 
\underbrace{\theta_{**} \cdots \theta_{**}\;}_n \;
\langle\bar{\psi} P^{r^1}_{l^1_0 l^1_1\dots l^1_{k_1}} \psi\rangle 
\cdots 
\langle\bar{\psi} P^{r^s}_{l^s_0 l^s_1\dots l^s_{k_s}} \psi\rangle \;
P^{r^0}_{l^0_0 l^0_1\dots l^0_{k_0}} \psi
\Big)^{(i)},
  \label{psi}
\\
P^{r^j}_{l^j_0 l_1^j\dots l_k^j} &=
\underbrace{\gamma^* \cdots \gamma^*}_{r^j} 
\underbrace{D_* \dots D_*\;}_{l_1^j}(F_{**}) \cdots 
\underbrace{D_* \dots D_*\;}_{l_{k_j}^j}(F_{**}) 
\underbrace{\tilde{D}_* \dots \tilde{D}_*\;}_{l_0^j} \;,
\nonumber
\end{align}
where $\sum_{j=0}^s (2 k_j + \sum_{h=0}^{k^j} l^j_h) = 2n {-}t {-}3s$
and $\sum_{j=0}^s r^j=4n{-}t{-}3s$. These conditions guarantee that
$\hat{\psi}^{(n)}{}'$ has the same power-counting dimension
($=\tfrac{3}{2})$ as $\psi$. All indices are summation indices.  We
have introduced the covariant derivative for fermions $\tilde{D}_\mu
\psi= \partial_\mu \psi- \mathrm{i} A_\mu \psi$, $m$ is the fermion
mass and $\gamma^\mu$ are the Dirac gamma matrices.  The quantity
$\langle\bar{\psi} P^{r}_{l_0 l_1\dots l_{k}} \psi\rangle$ is a (gauge
invariant) function on space-time obtained by taking the trace in
spinor and colour space, without space-time integration.

In the same way as in (\ref{ginv}), $\hat{\psi}^{(n)}{}'$ is a
solution of the gauge-equivalence (\ref{ge}) if $\hat{\psi}^{(n)}$ is:
\begin{align}
\delta_{\hat{\lambda}} \hat{\psi}^{(n)}{}' =   
\delta_{\hat{\lambda}} \hat{\psi}^{(n)} + \mathrm{i} \lambda  \Psi^{(n)} =
\delta_{\lambda} \hat{\psi}^{(n)}{}'\qquad 
\text{up to order $n$.}
\end{align}

The Seiberg-Witten map for the adjoint spinor $\bar{\psi}=\psi^\dag
\gamma^0$ is simply obtained by Hermitean conjugation, using
$\gamma^\mu{}^\dag \gamma^0=\gamma^0\gamma^\mu$: A term 
\[
\kappa\; \Pi \gamma^{\mu_1} \cdots \gamma^{\mu_r} 
P^{0}_{0l^0_1\dots l^0_{k_0}}
\tilde{D}_{\nu_1} \dots 
\tilde{D}_{\nu_l}\psi
\]
in $\Psi^{(n)}$, where $\Pi$ contains all saturated fermions
$\langle\bar{\psi} P^{r}_{l_0 l_1\dots l_{k}} \psi\rangle$, is
transformed into
\[
\bar{\kappa} \;
(\tilde{D}_{\nu_1}^\dag \dots 
\tilde{D}_{\nu_l}^\dag)(\bar{\psi})\,
\gamma^{\mu_r} \cdots \gamma^{\mu_1} P^{0}_{0 l^0_{k_0}\dots l^0_1}
\bar{\Pi} 
\]
in $\bar{\Psi}^{(n)}$, where $\tilde{D}_\nu^\dag \bar{\psi} = \partial_\nu
\bar{\psi} +\mathrm{i} \bar{\psi} A_\nu$. 

Then, the noncommutative Dirac action 
\begin{align}
\label{SF}
\hat{\Sigma}_D = \int
d^4x\;\Big(\langle\hat{\bar{\psi}} 
(\mathrm{i}\gamma^\mu \partial_\mu -m)\hat{\psi}\rangle + 
\langle \hat{\bar{\psi}} \gamma^\mu \hat{A}_\mu \star 
\hat{\psi}\rangle\Big)
\end{align}
 gives after
Seiberg-Witten map the real-valued gauge invariant fermionic action
\begin{align}
  \label{AF}
\Sigma_D^{(n)}{}' =\Sigma_D^{(n)} + \int d^4x\;\Big(
\big\langle \bar{\psi} \big(\gamma^\mu(\mathrm{i}\partial_\mu +A_\mu)-m\big)
\Psi^{(n)}\big\rangle + 
\big\langle \bar{\Psi}^{(n)}\big(\gamma^\mu(\mathrm{i}\partial_\mu 
+A_\mu)-m\big)\psi \big\rangle \Big) ~.
\end{align}
The part $\Sigma_D^{(n)}{}' -\Sigma_D^{(n)}$ is due to (\ref{psi}) a
real-valued gauge invariant integrated field polynomial of
power-counting dimension 0 with at least two fermions.  Such terms
will also come from the action $\Sigma_D^{(n)}$, which leads
effectively to a shift of $\tilde{\kappa}_i^{(n)}$. However, this
generates only a subset of all gauge invariant fermionic actions
\cite{QED}. The hope is that (assuming again an invariant
renormalization scheme) the (divergent) 1PI Green's functions are
precisely of the form (\ref{AF}). As for the $N$-point photon
functions with $N\geq 3$, the Ward identity gives no further information.

Assuming it is possible to prove that divergent 1PI Green's functions
are of the form (\ref{AF}), let us show that the Seiberg-Witten map
(\ref{psi}) for fermions can be extended to order $n{+}1$.
This goes as in the bosonic case via Taylor expansion and the
differential equation implementing the gauge-equivalence:
\begin{align}
\label{dglf}
\frac{\partial \hat{\psi}}{\partial \theta^{\alpha\beta}}
= -\frac{1}{8} \Big(2 \hat{A}_\alpha \star \partial_\beta \hat{\psi}
- \partial_\alpha \hat{A}_\beta \star
\hat{\psi}\Big) 
+\frac{1}{8} \Big(2 \hat{A}_\beta \star \partial_\alpha \hat{\psi}
- \partial_\beta \hat{A}_\alpha \star \hat{\psi}\Big)~.
\end{align}
Then,
\begin{align}
\hat{\psi}^{(n+1)}{}' & = \psi - \frac{1}{4} 
\sum_{k=1}^{n+1} \frac{1}{k!}
\theta^{\alpha_1\beta_1} 
\cdots \theta^{\alpha_k\beta_k} 
\Big(\frac{\partial^{k-1}}{\partial 
  \theta^{\alpha_2\beta_2} \dots
  \theta^{\alpha_k\beta_k}}
\Big(2 \hat{A}^{(n)}_{\alpha_1}{}'\! \star \partial_{\beta_1} 
\hat{\psi}^{(n)}{}' 
{-} \partial_{\alpha_1} \hat{A}^{(n)}_{\beta_1}{}' \! \star 
\hat{\psi}^{(n)}{}' \Big)\!
\Big)_{\theta=0} 
\nonumber
\\
  \label{n+1F}
& + \Psi^{(n+1)}_\mu
\end{align}
is the required solution of the gauge-equivalence at order $n{+}1$ in
$\theta$. Again, $\hat{\psi}^{(n+1)}{}'$ is a complicated nonlinear
function of $\kappa^{(j)}_i$ and $\tilde{\kappa}^{(j)}_i$ for $j\leq
n$. 

\section{Discussion}

We have proved renormalizability of the photon self-energy in
noncommutative QED to all orders in perturbation theory. This is the
first example of a renormalizable Green's function in a noncommutative
gauge theory.  After the classification of diseases of noncommutative
QFTs by Chepelev and Roiban \cite{Chepelev:2001hm} there remained not
much hope that this could be achieved beyond one-loop.

The alternative approach via the Seiberg-Witten map
\cite{Seiberg:1999vs} introduces an infinite number of
non-renormalizable vertices with unbounded power-counting degree of
divergence into the game. It is therefore surprising that at least for
the photon self-energy such bad divergences can be treated.
Fortunately the Seiberg-Witten map is a friendly monster which for
each problem in a given order provides a cure in the same order (by
shifting the mess to the next order, etc).

In this way we have achieved renormalization of a Green's function in
a gauge theory with an external field of negative power-counting
dimension -- a model with infinitely many vertices. The point is that
via the Seiberg-Witten map all these vertices can be summed up to an
action as simple as (\ref{Sig}). There exist closed formulae for the
Seiberg-Witten map to all orders in $\theta$, see \cite{Jurco:2001my}
and references therein. In \cite{Jurco:2001my} there was also given an
abstract definition of the freedom in the Seiberg-Witten map which
should contain the field redefinitions we used to show
renormalizability of the photon self-energy. It should be stressed
however that only concrete loop calculations such as done in
\cite{Bichl:2001nf} can determine the parametrization (\ref{renorm})
which renormalizes the photon self-energy.

Of course the renormalizability proof should be extended to other
Green's functions than the photon self-energy. This is an open
problem, but it is plausible now that noncommutative QED is
renormalizable. Indeed, the photon self-energy contains (at high
enough loop order) graphs of any other Green's function as
subdivergences. These subdivergences assumed to be treated according
to the forest formula, we know that the overall divergence of the
photon self-energy is renormalizable. The open question is whether the
Green's functions of these subdivergences can give rise to
counterterms incompatible with the noncommutative action after field
reparametrizations.

A main goal is of course to formulate a renormalizable noncommutative
version of the standard model. In this respect we stress that in
$\theta$-deformed QED there is only one place for a coupling constant
-- namely in front of the photon action.  It is therefore not possible
to have fermions of different electric charge \cite{Hayakawa:1999zf}.
This is not a problem because in noncommutative geometry a part of the
electric charge of the quarks comes from the colour sector
\cite{Connes:1996gi}.

One of the basic principles of renormalization is the independence of
the specific way one treats the problems. How can we understand then
the UV/IR problem \cite{Minwalla:2000px,Matusis:2000jf} which plagues
the $\theta$-undeformed approach and which is completely absent in the
Seiberg-Witten framework? We believe that the UV/IR mixing is not
really there, it is a non-perturbative artefact absent in perturbation
theory -- and thus should be treated by non-perturbative techniques as
suggested in \cite{Chepelev:2001hm}. Let us consider the integral
\[
I=\int d^4 k \; \frac{\mathrm{e}^{\mathrm{i} \tilde{p}_\mu
    k^\mu}}{k^2}~,
\qquad \tilde{p}_\mu := \theta_{\mu\nu} p^\nu~,
\]
which is part of the tadpole graph in noncommutative Maxwell
theory. The standard integration methods agree in the following
(finite!) answer:
\begin{align}
  \label{nonp}
I=\int d^4 k \; \frac{\mathrm{e}^{\mathrm{i} \tilde{p}_\mu k^\mu}}{k^2}
= \frac{4\pi^2}{\tilde{p}^\mu \tilde{p}_\mu}~.  
\end{align}
This $(1/p^2)$ behaviour is the origin of all infrared problems.
On the other hand, expanding the exponential we produce at first
sight divergences of arbitrary degree:
\[
I=\int d^4 k \; \frac{\sum_{n=0}^\infty \frac{1}{n!} 
(\mathrm{i} \tilde{p}_\mu k^\mu)^n}{k^2}~.
\]
Exchanging the sum and the integration, the integral of any term in the
series is scale-independent and IR well-behaved -- 
and as such zero in all standard renormalization schemes:
\begin{align}
  \label{p}
I=\sum_{n=0}^\infty \frac{1}{n!} \int d^4 k \; \frac{
(\mathrm{i} \tilde{p}_\mu k^\mu)^n}{k^2} =0~.  
\end{align}
The infrared problem disappeared.  There is no contradiction between
(\ref{nonp}) and (\ref{p}) because the integral is clearly not
absolutely convergent so that exchanging sum and integration is
dangerous. Which one of (\ref{nonp}) and (\ref{p}) is correct? There
are good reasons to believe that the $\theta$-perturbative result
(\ref{p}) should be preferred -- it leads to a renormalizable photon
self-energy. In some sense this can be regarded as a normal ordering
in noncommutative renormalization: First the integrals must be
performed, then the sums.  This eliminates the infrared singularities.

\section*{Acknowledgement}

First of all we would like to thank Elisabeth Kraus for bringing to
our attention the special role of field redefinitions, and for giving
valuable comments on an early version of this paper. Furthermore, we
would like to thank Karl Landsteiner, Dieter Maison, Stefan Schraml,
Peter Schupp, Klaus Sibold, Raymond Stora and Julius Wess for fruitful
discussions.

\end{document}